\documentstyle[prl,multicol,epsf,epsfig,aps]{revtex}
\newcommand{\BEQ}{\begin{equation}}
\newcommand{\EEQ}{\end{equation}}
\newcommand{\BEA}{\begin{eqnarray}}
\newcommand{\EEA}{\end{eqnarray}}

\renewcommand{\d}{{\rm d}}

\newcommand{\half}{\frac{1}{2}}

\def\dbarrm {{\mathchar'26\mkern-11mu{\rm d}}}                       %
\begin{document} 
\draft
\title
{Extraction of work from a single thermal bath in the quantum regime} 
\author{A.E. Allahverdyan$^{1,2,4)}$ and Th.M. Nieuwenhuizen$^{3)}$}
\address{$^{1)}$ CEA/Saclay, Service de Physique Theorique, 
F-91191 Gif-sur-Yvette Cedex, France, \\
$^{2)}$ Institute for Theoretical Physics, 
$^{3)}$ Department of Physics and Astronomy,\\ 
University of Amsterdam,
Valckenierstraat 65, 1018 XE Amsterdam, The Netherlands 
\\ $^{4)}$Yerevan Physics Institute,
Alikhanian Brothers St. 2, Yerevan 375036, Armenia }
\date{April 1, 2000}
\maketitle

\begin{abstract}
The stationary state of a quantum particle strongly coupled 
to a quantum thermal bath is known to be non-gibbsian, 
due to entanglement with the bath. 
For harmonic potentials, where the system can be described by effective 
temperatures, thermodynamic relations are shown to take a 
generalized Gibbsian form, that may violate the Clausius inequality.
For the weakly-anharmonic case a Fokker-Planck type description is 
constructed. It is shown that then work can be extracted from the 
bath by cyclic variation of a parameter. These apparent violations 
of the second law are the consequence of quantum coherence in the 
presence of the slightly off-equilibrium nature of the bath.
\end{abstract}

\begin{multicols}{2}

The laws of thermodynamics are at the basis of our
understanding of nature, so we all expect them to govern 
also systems coupled to a bath in the quantum regime. 
However, recently thought-provoking claims were made about
a violation of Thomson's formulation of the second law 
(the impossibility to do work periodically without loosing
heat)\cite{capek} and even about a perpetuum  mobile acting 
in an inhomogeneous superconducting ring~\cite{nikulov}.

Most of our thermodynamic understanding is based on 
the Gibbs distribution. 
The laws of equilibrium thermodynamics apply equally well
to closed classical and quantum systems, as to open 
classical subsystems \cite{klim}.
The setting for the classical case is well known:
Under general statistical conditions \cite{klim,gardiner,risken,weiss}
one derives a Langevin equation.
The corresponding probability distribution is described by the
Fokker-Planck
equation, and it converges in time to the Gibbs distribution. 

Much less is known about the quantum Langevin equation
\cite{klim,gardiner,weiss,sen,ul}. The stationary distribution
has only been obtained for the harmonic potential. It
depends explicitly on the damping constant and becomes Gibbsian
only in the limit of weak damping \cite{klim,gardiner}, thus
preventing the applicability of equilibrium thermodynamics.
Entanglement is the very reason of this crucial difference, 
as subsystems are necessarily in a mixed state. 

In the present paper we examine the standard model for quantum 
Brownian motion, the so-called Caldeira-Leggett model
~\cite{caldeira}, see eq. (\ref{hamiltonian}). 
Hereto we employ methods developed recently for glasses \cite{1}.
For a particle in harmonic potential we define effective temperatures,
and put the thermodynamic relations in a generalized Gibbsian form. 
For weakly anharmonic confining potentials 
 Fokker-Planck equations will be constructed, which allow 
to obtain the stationary distribution 
and elucidate important aspects of nonstationary properties.

We shall provide a nontrivial thermodynamic interpretation for
the  relaxation towards the steady non-gibbsian state and
for the slow change of a system parameter.
Our main results are rather dramatic, apparently contradicting 
the second law: We show that the Clausius inequality 
$\dbarrm Q\le T\d S$ can be violated, and that it is 
even possible to extract work from the bath by
cyclic variations of a parameter (``perpetuum mobile'').
The physical cause for this appalling behavior
will be traced back to quantum coherence in the presence of the 
near-equilibrium bath.

The {\it  Quantum Langevin equation}
is derived from the exact Hamiltonian description of a 
particle and a thermal bath, 
when tracing out the degrees of freedom of the bath.
For $t<0$ the particle and bath do not interact, and
the whole system is described by a density matrix
$\rho _0=\rho _b\otimes\rho _s$, where 
$\rho _b$ is the Gibbs distribution for 
the bath, and $\rho _s$ describes the state of the particle.
At $t=0$ a linear coupling is switched on instantaneously, and
the total Hamiltonian reads for $t>0$~\cite{weiss}
\begin{eqnarray}
\label{hamiltonian}
&&{\cal H}=\frac{p^2}{2m}+V(x)+\sum_{i}\left [
\frac{p_i^2}{2m_i}+\frac{m_i\omega_i^2}{2}
(x_i - \frac{c_i\,x}{m_i\omega_i^2} )^2
\right]
\end{eqnarray}
where $p$, $p_i$, $x$, $x_i$, and $m$, $m_i$ 
are, respectively,
the momentum and coordinate operators and the masses
of the particle and the modes of the bath. 
The latter have a constant frequency gap $\Delta$,
so $\omega _i=i\,\Delta$. For the couplings 
we choose the Drude-Ullersma spectrum~\cite{weiss,ul,haake}
\BEQ 
c_i=\sqrt{\frac{2\gamma m_i\omega_i^2\Delta}{\pi}\,
\frac{\Gamma^2}{\omega_i^2+\Gamma^2}},
\EEQ 
where $\gamma$ is the damping constant, quantifying the
stength of interaction, and $\Gamma $ is the cut-off 
frequency of the interaction with the bath.
The thermodynamic limit is taken for the bath by sending 
$\Delta\to 0$. Notice that this creates an infinite timescale
$1/\Delta$, implying that in the remaining approach the 
limit of ``large times'' always means the 
state where time is still much less than $1/\Delta$.
 
For fast switching on the interaction, an amount of work
\BEQ
\label{switch}
{\cal W}_0={\rm tr}\{\rho _0[ {\cal H}(0^+)-{\cal H}(0^-)]\}
=\frac{\gamma\Gamma}{2}{\rm tr}(\rho _sx^2)
\EEQ
has to be supplied to the system.
The large internal energy of the whole system 
$U=\langle p^2\rangle_0/2m+\langle V\rangle_0
+\pi^2T^2/(6\hbar\Delta)$$\sim $$1/\Delta$ is 
increased by the finite amount ${\cal W}_0$. 
This brings the system slightly out of equilibrium, 
since the von Neumann entropy ${\rm tr}(-\rho _0\ln \rho _0)$ is 
exactly conserved under the switching. Notice that the switching energy
${\cal W}_0=\half{\gamma\Gamma}\langle x^2\rangle_0$ is 
purely a classical effect.

The resulting quantum Langevin equation reads \cite{gardiner,sen}
\begin{eqnarray}
\label{01}
\dot{p}+\frac{\gamma\Gamma}{m}\int _0^t\d t^{\prime}
e^{-\Gamma (t-t^{\prime})}{p(t^{\prime})}
 +V^{\prime}(x)=\eta (t)
\end{eqnarray}
with  Gaussian noise having $\langle\eta\rangle=0$ and anticommutator
\begin{eqnarray}
\label{Kt=}
K(t)=\frac{\langle\{\eta (t),\eta (0) \}\rangle}{2}
=\int  \frac{\d \omega}{2\pi} \frac{\gamma
\hbar\omega \coth (\half \beta\hbar\omega)\,e^{i\omega t}}
{1+(\omega /\Gamma )^2}
\end{eqnarray}
The connection between properties of the noise and the friction kernel
is the consequence of quantum fluctuation-dissipation theorem 
\cite{klim,weiss}.
Eq.~(\ref{01}) with physically suitable forms of the potential
and friction describes a rich variety of physical phenomena,
such as Josephson junctions \cite{deti,van},
processes in plasma and condensed matter \cite{klim,gardiner,ul},
interaction of atoms with black-body radiation \cite{klim},
and the Lamb shift of an electron \cite{klim}.

We shall restrict ourselves to the {\it quasi-Ohmic} case, where $\Gamma$
is much larger than other characteristic times.
For $t\gg 1/\Gamma $ we may expand the memory kernel  in Eq.~(\ref{01})
\begin{eqnarray}
\label{1.1}
\Gamma\int _0^t\d t^{\prime}
 e^{-\Gamma (t-t^{\prime})}p(t^{\prime})
 =p(t)-\frac{\dot{p}(t)}{\Gamma }
\end{eqnarray}
Compared to the classical white noise case,
one can thus adjust only the noise, see eq. (\ref{Kt=}), 
while keeping the friction instantaneous. 
The same conclusion was reached in refs.~\cite{dekker,hu} for 
the exactly solvable harmonic potential.

{\it Work and heat.}
Let us explain the ingredients for the thermodynamic description of
the Brownian particle having 
Hamiltonian $H=p^2/2m+V(x)$ and Wigner function $W(p,x,t)$.
A change with time of the mean energy 
$U=\langle H\rangle \equiv \int \d p\d x\, W H$ 
is considered when slowly varying 
a system parameter $\alpha $, such as $m$ or $V''(0)$,
according to a prescribed trajectory $\alpha (t)$:
\begin{equation}
\label{dE}
\d U\equiv
\d \langle H\rangle=\langle H\d \ln W\rangle+\langle\d H\rangle
\end{equation}
The last term is the averaged mechanical work
$\dbarrm {\cal W}$ produced by external sources \cite{klim,sasa}.
For a variation of an intrinsic parameter 
(for which $\d H=\d {\cal H}$),
$\dbarrm {\cal W}$ is the change of the total
(particle$+$bath) mean energy.
The first term in r.h.s. is
due to the statistical redistribution of the 
phase space. We shall identify it with the change in heat 
$\dbarrm {\cal Q}=\langle H\d \ln W\rangle$, so Eq. (\ref{dE})
is just the first law.

In the harmonic case $ V(x)=\half ax^2$, Eqs.~(\ref{01},\ref{1.1})
can be integrated directly, 
see e.g.~\cite{weiss,dekker,haake,hu}. 
In particular, the stationary Wigner distribution reads
\begin{eqnarray}
\label{ole77}
W_{s}(p,x)=
\frac{1}{2\pi}\sqrt{\frac{a}{mT_pT_x}}\,\,
\exp{[-\frac{p^2}{2mT_p}-\frac{ax^2}{2T_x}]}.
\end{eqnarray}
The effective temperatures $T_p$ and $T_x$ approach $T$ in the 
classical limit, and have for large damping, $\gamma^2\gg am$ 
the following values at $T\to 0$:
\begin{eqnarray}
\label{TT0}
T_p=\frac{\hbar \gamma }{\pi m}\ln\frac{\Gamma m}{\gamma}+
\frac{\hbar a}{\pi \gamma},\, 
~T_x=\frac{\hbar a}{\pi\gamma}\ln\frac{\gamma^2}{am}
\end{eqnarray}
The brownian particle has semiclassical behavior due to  
its interaction with the bath, and entropy $S=S_p+S_x$ with 
$S_p=\half\ln emT_p/\hbar$, $S_x=\half\ln eT_x/a\hbar$.

For an adiabatic variation of a parameter $\alpha $ the 
situation 
is still described by Eq.~(\ref{ole77}) with $\alpha =\alpha (t)$.
One can check for $U$ and the free energy $F=U-T_pS_p-T_xS_x$
\begin{eqnarray}
\label{asala2}
&&\d U=\dbarrm Q_{ad}+\dbarrm {\cal W}_{ad}
=T_p\d S_p+T_x\d S_x+\dbarrm {\cal W}_{ad}, \\
\label{asala1}
&&\d F=- S_x\d T_x- S_p\d T_p+\dbarrm {\cal W}_{ad}.
\end{eqnarray}
These generalized thermodynamical relations
are in close analogy with those proposed 
for  glassy systems \cite{1}.

To demonstrate a violation of the fundamental
Clausius inequality $\d Q\le T\d S$ we consider a slow variation 
of the mass $m$ at $T\to 0$. Using Eqs.~(\ref{ole77}, \ref{TT0},
\ref{asala2}) one gets
\begin{eqnarray}
\label{dEEm}
\dbarrm {\cal W}_{ad}=-T_p\frac{\d m}{2m},\,
~\dbarrm {\cal Q}_{ad}=\d U-\dbarrm {\cal W}_{ad}
=\frac{\hbar \gamma }{m^2\pi  }\frac{\d m}{2}
\end{eqnarray}
Thus, there is the transfer of heat even for $T=0$,
and the Clausius inequality is violated when $\dbarrm Q>0$, 
i.e. for $\d m >0$. The latter also happens when
varying $a$.

{\it The Fokker-Planck equation}.
Except for the solvable harmonic potential, Eq.~(\ref{01}) 
is hardly tractable. Another approach  
having started from (\ref{01}), goes further to 
the weak-coupling (small $\gamma$) limit described by a 
Markovian master-equation
\cite{gardiner}. 
In the opposite, strong-coupling limit 
one can consider the terms in (\ref{01}) as c-numbers, 
but with the Gaussian quantum noise \cite{deti,van}.
The correspondence with the underlying quantum problem is established
through the Wigner function, i.e.,  
$\langle \delta (p(t)-p)\delta (x(t)-x)\rangle=W(p,x,t) $.
Since this approach is 
still exact for the harmonic case, one of the conditions of its validity 
involves a characteristic scale $L$ where the non-linearity
remains small: $ L\gg\sqrt{\hbar/\gamma}$ \cite{deti}.
Based on this condition 
we have derived a closed equation for the Wigner function.
Here we only give the final result, 
while details will be presented elsewhere~\cite{ANprep},
\begin{eqnarray}
\label{ko1}
&&\frac{\partial W(x,p,t)}{\partial t}=
-\frac{p}{m}\frac{\partial W}{\partial x}+\frac{\partial}{\partial p}
\left ([\frac{\gamma}{m}p+V^{\prime}(x)]W \right )
\nonumber\\
&&+\frac{\partial ^2}{\partial p\partial x}\left (D_{xp}(x,t)W\right )
+\gamma D_{pp}(x,t)
\frac{\partial ^2W}{\partial p^2},
\end{eqnarray}
For our purposes it is enough to indicate the stationary values
of the diffusion coefficients $D_{xp}$ and ${D}_{pp}$, 
\begin{eqnarray}
\label{15}
\label{Dpp}
&&D_{pp}(x)=
\int^{\infty}_0\frac{\d \omega}{\pi m}
\frac{\bar{K}(\omega)\omega^2}{(\omega^2+\omega_1^2)(\omega^2+\omega_2^2)}
\nonumber\\
\label{Dxp}
&&D_{xp}(x)=
\int^{\infty}_0\frac{\d \omega}{\pi m^2}
\frac{\bar{K}(\omega)(V^{\prime\prime}(x)-m\omega^2)}
{(\omega^2+\omega_1^2)(\omega^2+\omega_2^2)}
\EEA
where $\bar{K}(\omega)$ is the spectrum of $K(t)$ in Eq. (\ref{Kt=})
and 
$\omega _{1,2}=[{\gamma}
\pm \sqrt{{\gamma^2}-4{m}V^{\prime\prime}(x)}\,]/(2m)$.
Equation (\ref{ko1}) was derived under the same assumptions as the
semiclassical Langevin equation itself. 
For ensuring the convergence of the diffusion coefficients 
(${\rm Re}(\omega _{1,2})\ge 0$) we will demand 
$V^{\prime \prime}\ge 0$ (local stability).
In the classical limit: $D_{xp}\to 0$, $D_{pp}\to T$, and Eq.~(\ref{ko1})
tends to usual Fokker-Planck equation \cite{gardiner,haake}. 
For the harmonic case $D_{xp}$ and $D_{pp}$ become space-independent,  
and Eq.~(\ref{ko1}) is in agreement with previous results 
\cite{dekker,haake,hu}, obtained for a more general type of 
the environment. 

In the physically interesting case of overdamped motion, 
where the characteristic times of the momenta $m/\gamma$ and
coordinate $\gamma /V^{\prime\prime}(x)$ are widely separated
$\gamma /V^{\prime\prime}(x)\gg m/\gamma $, a solution
of the Fokker-Planck equation for $t\gg m/\gamma$ can be presented as
\begin{equation}
\label{ole44}
W(p,x,t)=\frac{\exp [-p^2/(2mD_{pp}(x,t))]}
{\sqrt{2\pi mD_{pp}(x,t)}}\,\,W(x,t)
\end{equation}
Here $W(x,t)$ is the solution of a reduced equation
\begin{eqnarray}
\label{ko2}
&&\partial_t W(x,t)+\frac{\partial J (x,t)}{\partial x}=0, \\
\label{J} && J=-\frac{1}{\gamma }V^{\prime}(x)W(x,t)
-\frac{1}{\gamma}\frac{\partial}{\partial x}[D(x,t)W(x,t)],\\
\label{D} && D(x,t)=D_{xp}(x,t)+D_{pp}(x,t).
\end{eqnarray}
It still contains the inhomogeneous, time-dependent diffusion coefficient. 
As we discussed, large, but finite $\Gamma\gg \omega _{1,2}$ 
is necessary only for the statistics of the momenta,
see Eq. (\ref{TT0}). 
Taking $\Gamma \to \infty$ in Eq.~(\ref{D}) we obtain 
\begin{eqnarray}
D(x)=\frac{\hbar \gamma V^{\prime\prime}(x)}{\pi m^2}
\int_0^{\infty}\d \omega 
\frac{\omega\coth\left(\half\beta\hbar\omega\right)}
{(\omega^2+\omega_1^2)(\omega^2+\omega_2^2)}
\label{rashrash}
\end{eqnarray}
In the stationary state one has
$J(x)=0$, which implies for the corresponding distribution
\begin{eqnarray}
\label{statd}
W_{s}(x)=\frac{e^{-\beta V_e(x)}}{Z};\,\,{V_e(x)}=T
\int_0^x\d y\frac{V^{\prime}(y)+D'(y)}{ D(y)} 
\end{eqnarray}
where $V_e$ is an effective potential $V_e$. Notice that
$W_{s}(x)$ is non-gibbsian. 
For the harmonic potential Eq. (\ref{ole77})
is recovered from this expression 
while the classical Gibbs distribution appears in the limit 
$\hbar\beta \to 0$.
Eqs.~(\ref{ole44}, \ref{statd}) show that the 
statistics of momenta is influenced by the 
coordinate, yielding its non-Maxwellian form.

In classical case universal (thermodynamical) 
properties of the relaxation are described by an
${\cal H}$-theorem, which is intimately connected with a 
formulation of the second law \cite{klim,risken}.
This theorem can be generalized in our case, at least
for times $t\gg m/\gamma$, where the momenta already 
came to the local equilibrium, and the relevant variable is $x$.
The ${\cal H}$-function is defined as 
\cite{klim,risken}
\begin{equation}
\label{HHH}
{\cal H}=\int \d x
W_1(x,t)\ln \frac{W_1(x,t)}{W_2(x,t)}\ge 0,
\end{equation}
where $W_{1,2}(x,t)$ are solutions of 
Eq.~(\ref{ko2}) corresponding to different initial 
conditions. Calculating $\dot{{\cal H}}$ from Eq.~(\ref{ko2}) one  
performs partial integrations to obtain 
\begin{eqnarray}
\label{2203}
\dot{{\cal H}}
=-\frac{1}{\gamma}\int \d x\,
W_1(x,t)D(x,t)
\left [\frac{\partial}{\partial x} \ln\frac{W_1(x,t)}{W_2(x,t)} 
\right ]^2
\end{eqnarray}
Since $D(x,t)>0$,
${\cal H}$ is a monotonically decreasing function, and attains
its minimum in the stationary state \cite{risken}. On the other
hand, ${\cal H}$ is limited from below by zero, 
so we conclude that 
all solutions of the Fokker-Planck equation (\ref{ko2}) 
converge with time to the stationary solution.
Let us define the current-less state 
$W_{s}(x,t)$, obtained from $W_{s}(x)$ by the substitution
$D(x)\to D(x,t)$.
Starting from Eq.~(\ref{2203}) one can show that the entropy 
of the coordinate sector obeys
\begin{equation}
\label{222}
\frac{\d S_x}{\d t}
=-\int \d x \dot{W}(x,t) \ln W(x,t)
=\frac{\d _e S_x}{\d t}+\frac{\d _i S_x}{\d t}.
\end{equation}
Here the quantity
\begin{eqnarray}
\frac{\d _eS_x}{\d t}=\int \d x\dot{W}(x,t)\ln \frac{1}{W_{s}(x,t)}
=\int \d x~J(x)\beta V^{\prime}_e(x,t)
\nonumber
\end{eqnarray}
can be interpreted as the flux of entropy. 
Indeed, $J(x)$ is the probability current and  $-\beta V_e'$ 
represents a force divided by temperature.
The second term in (\ref{222}),
\begin{eqnarray}
\label{o222}
\frac{\d _iS_x}{\d t}=\frac{1}{\gamma}\int \d x W(x,t)D(x,t)\left[
\frac{\partial}{\partial x}\ln\frac{W(x,t)}{W_{s}(x,t)}\right ]^2
\nonumber
\end{eqnarray}
is the {\it entropy production} during the relaxation. 
This quantity is strictly positive out of the steady state, 
and becomes zero in the long-time limit.
In the classical case Eq.~(\ref{222}) leads to the well-known relation
$T\d S=\d {\cal Q}+T\d _iS$, since $W_{s}(x,t)$ becomes time-independent
and coincides with the Gibbs distribution. 
However, in the general quantum case the relation between heat 
($\dbarrm {\cal Q}$) and the flux of the statistical entropy 
($\dbarrm {\cal Q}=T\d _eS$) appears to be broken.
A similar relation can, though, still be recovered in the harmonic 
case, where the diffusion constant $D$ does not depend on the 
coordinate. Applying Eq.~(\ref{222}) we find
$T_p\d S_p+T_x \d S_x=\d U +\dbarrm \Pi$,
where $T_p(t)=D_{pp}(t)$ and $T_x(t)=D(t)$ are
the time-dependent effective temperatures, and 
$\dbarrm \Pi=T_x \d _i S_x\ge 0$
is the energy dissipated during the relaxation.

Returning to the case of varying a system parameter, we are now
interested in the first non-adiabatic correction to the 
stationary distribution, arising when the time of the 
variation is large, but finite.
Since the dynamics is in the overdamped regime, 
the main correction comes from the deviation of  $W(x,t)$
(and not the full $W(p,x,t)$) from its stationary form.
The variation starts at $t=t_i$ (the particle already reached
its stationary state), and ends at $t=t_f$.
We shall assume a ``smooth'' start of the variation,
i.e., $\dot{\alpha }(t_i)=0$.
It can be checked directly from Eq.~(\ref{ko2}) that
$W$ now reads for small $\dot\alpha$
\begin{eqnarray}
\label{6d6}
&& W(x,t)=W_{s}(x,\alpha )\left [1+ \frac{\gamma}{T} \dot{ \alpha}
\left (B(x,t)-\langle B\rangle\right )\right ], \\
\label{6d6d6}
&&B(x,t)
=\int^x_{-\infty}\frac{\d y}{D(y,\alpha)W_{s}(y,\alpha)}
{\cal A}(y,\frac{ \partial  V_e}{\partial\alpha}),
\label{55}\\
\label{urra1}
&&{\cal A}(y,f)=
\int^{\infty}_{y}\d z W_{s}(z,\alpha)(f(z)-\langle f\rangle),
\end{eqnarray}
where $\alpha=\alpha(t)$ and the average is taken 
w.r.t. $W_{s}(x,\alpha)$.
The work finally becomes 
$\dbarrm{\cal W}=\dbarrm{\cal W}_{ad}\,+\,\dbarrm\Pi,$ with
\begin{eqnarray}\label{urra}
&&\dbarrm {\cal W} _{ad}=\dot \alpha\,\d t\, 
\int \d x\d p W_{s}(p,x, \alpha )\frac{\partial H(p,x,\alpha )}
{\partial \alpha },
\label{1.5law}
\\
&&\label{dPi=int}
\dbarrm\Pi=\frac{\gamma}{T}
\dot{\alpha}^2\,\d t\,
\int \frac{\d x~ {\cal A}\left ( x,
 \partial _{\alpha}  V
\right ){\cal A}\left ( x,  \partial _{\alpha}  V_e
\right )    }{D(x,\alpha )
W_{s}(x,\alpha )}
\label{kaban}
\end{eqnarray}
In the Gibbsian case at large $T$, where $V_e\to V$, one recovers
${\cal W}_{ad}=U-TS$ and the known result for
$\dbarrm \Pi \ge 0$ \cite{klim,sasa}. 
For low $T$  $\dbarrm \Pi$ remains finite due to the explicit 
factor $T$ in $V_e$, see (\ref{statd}).
We stress that $\dbarrm \Pi$ is always relevant
for cyclic processes $\alpha (t_i)=\alpha (t_f)$,
where $\Delta {\cal W}_{ad}=0$. 

In the harmonic case $\alpha$ can stand for the spring constant $a$.
It is then  straightforward to see that $\dbarrm\Pi\sim
\partial(a/T_x)/\partial a$. As expected, this is always positive.

For the oscillator with $V(x)=ax^2/2+gx^4/12$ 
with small $g$ the anharmonicity is displayed at the
scale $L=\sqrt{a/g}\gg \langle |x|\rangle$. 
We shall now investigate the non-adiabatic correction
to the work caused by temporal variation of $L$.
For $T\to 0$ we additionally take the limit of large $\gamma $ 
in Eqs.~(\ref{statd}, \ref{rashrash}), which yields
\begin{eqnarray}
\label{v3}
&&\beta V_e(x)\to\frac{\gamma\pi}{2\hbar\ln \gamma}\left [
\frac{x^2}{6}+\frac{L^2}{3}\ln (1+\frac{x^2}{L^2})
\right ].
\end{eqnarray}
Although $\partial _L  V(x) = -[a/(6L^3)]x^4$
is negative, $\partial _L V_e(x)\approx
+[\gamma\pi T/(6\hbar L^3 \ln\gamma)]\,x^4$
is positive.  Therefore $\dbarrm\Pi$ is negative. 
For small $g$ we may calculate the
integral in Eq.~(\ref{dPi=int}) in the harmonic approximation, to get
\begin{equation}\label{v9}
\Delta\Pi =-\frac{7}{3}\gamma{\langle x^2\rangle ^3}
\int^{t_f}_{t_i}
\d t \,\frac{\dot L^2(t)}{{L^6(t)}},
\end{equation}
where $\langle x^2\rangle =(2\hbar\ln \gamma) /(\pi\gamma)$ 
is the dispersion in this approximation. 
The possibility to extract this energy from the bath
is due to its non-equilibrium state, which is ensured by 
the energy (\ref{switch}) supplied in the switching.

In conclusion, we have considered a brownian quantum particle 
strongly interacting with a quantum thermal bath. 
The non-gibbsian statistics of the particle is completely described
by Fokker-Planck equations (\ref{ko1}, \ref{ko2}). 
An ${\cal H}$-theorem is formulated in eqs. (\ref{2203}), (\ref{222}). 
For the harmonic potential generalized Gibbsian relations can be 
constructed in terms of effective temperatures (\ref{ole77}), 
(\ref{TT0}), as happens also in glassy systems~\cite{1}.
Two formulations of the second law, namely the Clausius 
inequality and the impossibility to extract work during cyclical
variations, can be apparently violated at low temperatures. 
One could thus speak of a ``perpetuum mobile of the second kind''.
We should mention, however, that the number of cycles can be
large, but not arbitrarily large.
As a result, the total amount of extractable work is modest
~\cite{ANprep}.

These violations of the second law are due 
to quantum coherence in the presence of the slightly 
off-equilibrium nature of the bath. We call them apparent 
violations, since, the standard requirements for a thermal bath 
not being fulfilled, thermodynamics just does not apply.
Let us stress that also in the classical regime the harmonic oscillator
bath is not in full equilibrium, but there the Gibbs distribution
saves the day and thermodynamics does apply.
Our results thus make clear that the characterization of the heat bath
should be given with care. If it thermalizes on a short timescale,
standard thermodynamics always applies. If it does not thermalize 
within the observation time, it acts as a ``mechanical'' part 
of the system, and thermodynamics need not have a say. 
Although it still applies in the classical case, 
we have discussed that it fails to do so in the quantum case.

Let us close by noting that in the harmonic case the
unequal effective temperatures do not cause heat currents
that equalize them.
This situation is reminiscent of the classical
paradox that atoms should radiate,
but, being in the quantum regime, they do not.

We thank H. Dekker for useful discussions. A.E. A 
is grateful for support by FOM (The Netherlands).


\end{multicols}
\end{document}